\documentstyle[twocolumn,aps,prb]{revtex}

\begin{document}
\draft
\title{Snake orbits and related magnetic edge states }
\author{J. Reijniers and F. M. Peeters \cite{peeters}}
\address{Departement Natuurkunde, Universiteit Antwerpen (UIA), \\
Universiteitsplein 1, B-2610 Antwerpen, Belgium}
\date{\today}
\maketitle

\begin{abstract}
We study the electron motion near magnetic field steps at which the strength
and/or sign of the magnetic field changes. The energy spectrum for such
systems is found and the electron states (bound and scattered) are compared
with their corresponding classical paths. Several classical properties as
the velocity parallel to the edge, the oscillation frequency perpendicular
to the edge and the extent of the states are compared with their quantum
mechanical counterpart. A class of magnetic edge states is found which do
not have a classical counterpart.
\end{abstract}


\section{Introduction}

The transport properties of a two-dimensional electron gas (2DEG) subjected
to a nonhomogeneous perpendicular magnetic field (periodically modulated or
not) have been the focus of a great deal of research in recent years. \cite
{peeters99}\ Current fabrication technologies permit to create
nonhomogeneous magnetic fields on a nanometer scale by deliberately shaping
or curving the 2DEG,\cite{foden94} or by integration of superconducting\cite
{smith94,geim97} or ferromagnetic materials\cite{dubonos99,kubrak99} on top
of the 2DEG. This will add a new functional dimension to the present
semiconductor technology and will open avenues for new physics and possible
applications.\cite{johnson97}

Theoretically, the effect of nonhomogeneous magnetic fields on a 2DEG have
been studied both in the ballistic and the diffusive regime. The resulting
perpendicular magnetic field can act as a scattering centre,\cite
{dubonos99,kubrak99,reijniers00} but can also bind electrons,\cite
{solimany95,sim98,reijniers99,kim99} and so influence the transport
properties of the 2DEG. In transport calculations one needs the electron
states, which are obtained by solving the Schr\"{o}dinger equation.

M\"{u}ller\cite{muller92} studied theoretically the single particle electron
states of a 2DEG in a wide quantum waveguide under the application of a
nonuniform magnetic field and showed that in the case of a magnetic field
modulation in one direction, transport properties also become one
dimensional and electron states propagate perpendicularly to the field
gradient.$\ $

Making use of this decoupling, the electron states for different
nonhomogeneous magnetic field profiles along one dimension were
investigated, i.e. for a periodically modulated magnetic field \cite
{peeters93bis,ibrahim95,zwerschke99}, for magnetic quantum steps, barriers
and wells in an infinite 2DEG \cite{peeters93,calvo93} and in a narrow
waveguide.\cite{gu97}

In this paper we consider an infinite 2DEG subjected to a step-like magnetic
field, i.e abruptly changing in magnitude or polarity at $x=0$, in one
dimension (taken to be the $x$-direction). Preliminary results were
presented in Ref.~\onlinecite{peeters99bis}. First the situation for two
opposite homogeneous magnetic fields with the same strength will be
considered.$\ $The classical trajectories correspond to {\em snake orbits}
and were already used in the seventies\cite{history} to describe electron
propagation parallel to the boundary between two magnetic domains.$\ $Back
then, one was interested in understanding the electron transport through
multi-domain ferromagnets and it turned out to be more convenient to work
with the classical trajectrories than with the corresponding electron
states, which allows one to use a semi-classical theory which reduces the
complexity of the theory considerably. We are interested in transport
through a 2DEG situated in a semiconductor in which the Fermi energy is
orders of magnitude smaller than in the metallic systems of Ref.%
~\onlinecite{history}. 

We will study thoroughly the quantum mechanics of such electron states in a
2DEG subjected to this step magnetic field profile, and we will compare them
with their classical counterpart. We will discuss the energy spectrum and
the corresponding electron states, and derive several properties.$\ $We will
show the existence of states wich have a velocity opposite to the expected
classical orbits. Additionally, we will show that adding a background
magnetic field modifies the spectrum and the states considerably.$\ $

The paper is organized as follows.$\ $In Sec.~II we present our theoretical
approach.$\ $In Sec.~III we calculate the energy spectrum, the wavefunctions
and their corresponding group velocity, and compare this with their quantum
mechanical counterpart.$\ $In Sec.~IV we study the influence of a background
magnetic field on the quantum mechanical and classical behaviour. In Sec.~V
we focus on the negative velocity state, and finally, in Sec.~VI, we
construct time dependent states, and interpret them classically for several
magnetic field profiles.$\ $

\section{Theoretical approach}

We consider a system of noninteracting electrons moving in the $xy$-plane in
the absence of any electric potentials.$\ $The electrons are subjected to a
magnetic field profile $\vec{B}=(0,0,B_{z}\left( x\right) )$.$\ $First, we
will study the electronic states near the edge of two magnetic fields with
opposite strength

\begin{equation}
B_{z}\left( x\right) =B_{0}\left[ 2\theta (x)-1\right] ,
\end{equation}
which is independent of the $y$-coordinate. Next, we will consider the
influence of a background magnetic field $B$ on these states, which results
in the magnetic field profile

\begin{equation}
B_{z}\left( x\right) =B_{0}\left[ 2\theta (x)-1\right] +B.
\end{equation}
In the following we will use $B^{l}=B_{z}\left( x<0\right) $ and $%
B^{r}=B_{z}\left( x>0\right) $, to denote respectively the magnetic field on
the left and the right hand side of the magnetic edge.\ 

The one-particle states in such a 2DEG are described by the Hamiltonian 
\begin{equation}
H=\frac{1}{2m_{e}}p_{x}^{2}+\frac{1}{2m_{e}}\left[ p_{y}-\frac{e}{c}A\left(
x\right) \right] ^{2}.
\end{equation}
Taking the vector potential in the Landau gauge, 
\begin{equation}
\vec{A}=(0,xB_{z}\left( x\right) ,0),
\end{equation}
we arrive at the following 2D Schr\"{o}dinger equation 
\[
\left\{ \frac{\partial ^{2}}{\partial x^{2}}+\left[ \frac{\partial }{%
\partial y}+ixB_{z}(x)\right] ^{2}+2E\right\} \psi (x,y)=0,
\]
where the magnetic field is expressed in $B_{0}$, all lengths are measured
in the magnetic length $l_{B}=\sqrt{\hbar c/eB_{0}}$, energy is measured in
units of $\hbar \omega _{c}$, with $\omega _{c}=eB_{0}/m_{e}c$ the cyclotron
frequency and the velocity is expressed in units of $l_{B}\omega _{c}$.$\ H$
and $p_{y}$ commute due to the special form of the gauge, and consequently
the wavefuntion becomes 
\begin{equation}
\psi (x,y)=\frac{1}{\sqrt{2\pi }}e^{-iky}\phi _{n,k}(x),
\label{eq:wavefunction}
\end{equation}
which reduces the problem to the solution of the 1D Schr\"{o}dinger equation 
\begin{equation}
\left[ -\frac{1}{2}\frac{d^{2}}{dx^{2}}+V_{k}(x)\right] \phi
_{n,k}(x)=E_{n,k}\phi _{n,k}(x),  \label{eq:1d_schrod}
\end{equation}
where it is the $k$-dependent effective potential 
\begin{equation}
V_{k}\left( x\right) =\frac{1}{2}\left[ xB_{z}\left( x\right) -k\right] ^{2},
\label{eq:potential}
\end{equation}
which contains the two dimensionality of the problem.$\ $We will solve Eq.~(%
\ref{eq:1d_schrod}) numerically by use of a discretization procedure.$\ $In
some limiting cases analytical results can be obtained.

\section{In the absence of a background magnetic field}

Let us first consider the case when no background magnetic field is present.$%
\ $The situation is then symmetric, and more easily to solve.$\ $The
effective potential for this case is shown in Fig.~\ref{fig:trajectories}%
(a)\ for $k=-2$ (dotted curve) and $k=2$ (solid curve).$\ $We notice from
Eq.~(\ref{eq:potential}) that this potential is built from two parabolas,
with minima situated at $x^{l}=-k$, and $x^{r}=k$, thus respectively on the
left and right hand side of the magnetic edge. The total potential has for $%
k>0$ two local minima respectively at $x=-k$ and $x=+k$, while for $k<0$ it
has only one minimum at $x=0$. Before we describe the energy spectrum of the
snake orbits and their corresponding properties, we first discuss the
limiting behaviour.

\subsection{Limiting behaviour for $k\rightarrow \pm \infty $}

For $k\rightarrow \infty $, the minima of the parabolas are situated far
from each other.$\ $The electrons are in the Landau states of two opposite
magnetic fields, one on the left, the other on the right, and they are not
interacting with each other.$\ $The electron wavefunctions are given by $%
\left\langle L|x\right\rangle =C_{m}H_{m}(x+k)e^{-\left( x+k\right) ^{2}/2}$%
, and $\left\langle R|x\right\rangle =C_{m}H_{m}(x-k)e^{-\left( x-k\right)
^{2}/2}$, respectively, where $H_{m}\left( x\right) $ is the Hermite
polynomial.$\ $For decreasing $k$ the parabolas shift towards each other,
and the electrons will start to ``feel'' each other.$\ $In terms of
wavefunctions, this results in a parabolic cylinder function $\phi (x)=D_{E-%
\frac{1}{2}}\left[ \sqrt{2}\left( x-k\right) \right] $ matched at $x=0$,
with the condition that $\frac{d}{d\alpha }D_{E-\frac{1}{2}}\left( \alpha
\right) |_{\alpha =-\sqrt{2}k}=0$ or $D_{E-\frac{1}{2}}(-\sqrt{2}k)=0$, for
the symmetric and the antisymmetric wavefunction, respectively. This leads
to a change in energy of the electron states, which can be understood as a
lifting of the degeneracy of the two original electron wavefunctions.\ The
energy can then be written as $E_{\pm }(k)=\left\langle L\right| H\left|
L\right\rangle \left( k\right) \pm \left\langle L\right| H\left|
R\right\rangle \left( k\right) =E\left( k\right) \pm \Delta E\left( k\right) 
$ with the corresponding wavefunctions $\left| \phi \right\rangle =\left|
R\right\rangle \mp \left| L\right\rangle $.$\ $One can see that the presence
of the second parabola results in two effects: (1) a decrease of$\ E\left(
k\right) $ due to the finite presence of the wave function in the other
parabola, and (2) a splitting of the energy level due to the overlap, i.e.
one level ($E_{+}$) shifts upwards, while the other ($E_{-}$) shifts down.$\ 
$For $k\rightarrow \infty $, this results in the following first-order
approximation to $E$ and $\Delta E$: 
\begin{mathletters}
\begin{eqnarray}
E_{m}\left( k\right)  &=&\frac{1}{2}+m-\frac{2^{m-1}}{m!\sqrt{\pi }}%
k^{2m-1}e^{-k^{2}},  \label{En(infinity)} \\
\Delta E_{m}\left( k\right)  &=&\frac{2^{m}}{m!\sqrt{\pi }}%
e^{-k^{2}}k^{2m+1}.  \label{dEn(infinity)}
\end{eqnarray}

In the other limit $k\rightarrow -\infty $, the effective potential can be
approximated by a triangular well $V\left( x\right) =k^{2}/2-kB_{0}x$.$\ $%
Solutions for this potential consist of Airy functions, again matched at $x=0
$ with the condition that $\phi ^{\prime }\left( 0\right) =0$ or $\phi
\left( 0\right) =0$ which results, respectively into the anti-symmetric
wavefunction $\phi _{2m}\left( x\right) =C_{2m}\left( k\right) \left( \left|
x\right| /x\right) Ai\left[ z_{Ai^{\prime },m+1}+\left( 2k\right)
^{1/3}\left| x\right| \right] $ and a symmetric one $\phi _{2m+1}\left(
x\right) =C_{2m+1}\left( k\right) Ai\left[ z_{Ai,m+1}+\left( 2k\right)
^{1/3}\left| x\right| \right] $, respectively with energy

\end{mathletters}
\begin{mathletters}
\begin{eqnarray}
E_{2m}\left( k\rightarrow -\infty \right)  &=&\frac{1}{2}\left[
k^{2}-z_{Ai^{\prime },m+1}\left( 2\left| k\right| \right) ^{2/3}\right] ,
\label{E2n(-infinity)} \\
E_{2m+1}\left( k\rightarrow -\infty \right)  &=&\frac{1}{2}\left[
k^{2}-z_{Ai,m+1}\left( 2\left| k\right| \right) ^{2/3}\right] ,
\label{E2n+1(-infinity)}
\end{eqnarray}
where $z_{Ai,n}$ ($=-2.338$, $-4.088$, $-5.521$, $...$, $-\left[ 3\pi \left(
4n-1\right) /8\right] ^{2/3}$) and $z_{Ai^{\prime },n}$ ($=-1.019,$ $-3.248,$
$-4,820,$ $...,$ $-\left[ 3\pi \left( 4n-3\right) /8\right] ^{2/3}$), denote
respectively the $n^{th}$ $\left( n=1,2,3,...,\infty \right) $ zero of the
Airy function and of its derivative.$\ $One can see that for increasing
negative $k$, the difference between the two energy branches increases,
which is to first order linear in $\left| k\right| $. Namely the more
negative $k$, the narrower the well, thus the more the energy levels are
shifted from each other.$\ $

\subsection{Spectrum and velocity}

Solving Eq.~(\ref{eq:1d_schrod}) numerically gives rise to the energy
spectrum shown (solid curves) in Fig.~\ref{fig:spectrum}(a).$\ $For $%
k=\infty $, we obtain the earlier mentioned Landau levels, which are
labelled with the quantum number $m$.$\ $Each level is twofold degenerate.$\ 
$For decreasing $k$, the degeneracy is lifted and they separate into two
different branches with eigenstates $\left| 2m\right\rangle $ and $\left|
2m+1\right\rangle $ and eigenvalues $E_{2m}$ and $E_{2m+1}$, and
corresponding quantum numbers $n=2m$ and $n=2m+1$. This quantum number $n$
does not only result from arranging the levels according to their lowest
energy, starting with $n=0$, but it also reflects the number of nodes of the
corresponding wavefunction.$\ $Notice that the levels have now a non-zero
derivative, i.e. electrons propage in the $y$-direction, and their group
velocity is given by $v\left( k\right) =-\partial E\left( k\right) /\partial
k$.$\ $(The minus sign appears here because in Eq.~(\ref{eq:wavefunction})
we took $k_{y}=-k$). This group velocity is plotted (solid curves)\ in Fig.~%
\ref{fig:velocity}(a) for the 6 lowest levels. For $k=\infty $ electrons are
in a Landau level, and consequently there is no net current in the $y$%
-direction.$\ $Decreasing $k$, results in a net current in the $y$%
-direction, which is positive for the upper branches $\left( 2m+1\right) $,
but is initially negative for the branches $\left( 2m\right) $.$\ $For more
negative values of $k$ it increases almost linearly with increasing $\left|
k\right| $, which becomes the first order analytical result $%
v_{m}(k\rightarrow \infty )=-k$, obtained by differentiating (\ref
{E2n(-infinity)}) and (\ref{E2n+1(-infinity)}).$\ $

\subsection{Classical picture}

The center of the classical orbit corresponds to a zero in the effective
potential.$\ $The energy spectrum can be divided up into three regions which
can be classicaly understood by the electron orbits drawn in Fig.~\ref
{fig:trajectories}(a).$\ $In region (A) the electrons move in closed orbits
either in the magnetic field on the left or on the right hand side. Since
its cyclotron radius is smaller than the distance to the magnetic field
discontinuity, they feel a homogeneous magnetic field. There is no net
velocity. In region (B) the cyclotron radius intersects the magnetic field
discontinuity slightly, i.e. in such a way that the moving electron and the
center of its orbit are on the same side. The electron is nevertheless able
to penetrate in the opposite magnetic field region, which results in a
(rather small) propagation in the $y$ direction $v_{y}>0$. For $k=0$ the
center of the orbit is exactly on the edge between the two opposite magnetic
fields.$\ $In region (C) the center is located in the opposite magnetic
field region of which the electron is moving in, resulting in a faster
propagation of the electron in the $y$-direction. These different regions
are also indicated in Fig.~\ref{fig:spectrum}(a).

We can also make a quantitative classical study of the velocity, starting
from the quantum mechanical energy spectrum.$\ $Since in classical mechanics
there is no quantization, we make use of the obtained quantum energy
spectrum in order to find the classical energy and thus the radius of the
cyclotron orbit.$\ $Classically, the energy is contained in the circular
velocity $v_{\varphi }$ through $E\left( k\right) =v_{\varphi }^{2}\left(
k\right) /2$. For any given quantum mechanical $E(k)$-value we obtain
classically the circular velocity $v_{\varphi }(k)=\sqrt{2E(k)}$. Now if we
consider $x_{0}=\pm k$ to be the center of the electron orbit, we can
calculate for every $k$-value the classical velocity $v_{y}(k)$, since we
also know $v_{\varphi }(k)$ and the cyclotron radius $R(k)=v_{\varphi }(k)$.$%
\ $Using geometric considerations, we obtain the following relation 
\end{mathletters}
\begin{equation}
v_{y}(k)=v_{\varphi }(k)\sqrt{1-[k/v_{\varphi }(k)]^{2}}/\arccos[k/v_{\varphi }(k)],  
\label{eq:clasvel1}
\end{equation}
which is shown in Fig.~\ref{fig:velocity}(a) by the dotted curves. Comparing this with its quantum mechanical counterpart, we notice that for $%
k<0$ values good agreement is found, but for $k>0$ there is a large
discrepancy.$\ $Moreover, one can see that negative velocities cannot exist
classically.$\ $

The critical $k$-value, $k^{\ast }$, for which no classical propagating
states can exist, i.e. the electron describes just a circular orbit in a
homogeneous magnetic field, has to be equal to the cyclotron radius $k^{\ast
}=R\left( k^{\ast }\right) =v_{\varphi }\left( k^{\ast }\right) =\pm \sqrt{%
2E\left( k^{\ast }\right) }$, which leads to the boundary drawn in Fig.~\ref
{fig:spectrum}(a) (dotted parabola).

\section{With a background magnetic field}

With a background magnetic field three different configurations: a) $%
0<B<B_{0}$, b) $B=B_{0}$, and c) $B_{0}<B$ must be considered.$\ $In the
following we will study the snake orbits in these configurations.$\ $

\subsection{$0<B<B_{0}$}

Applying a background magnetic field $0<B<B_{0}$, results in a situation
which is very similar to the previous one.$\ $Again the two magnetic fields
have opposite sign, but in this case they also have a different strength,
i.e. $B^{l}=-B^{r}/p$.$\ $Again we can calculate analytically the correction
to the energy in the limit $k\rightarrow \infty $.$\ $For an electron on the
right hand side in the $m^{th\text{ }}$Landau state of a magnetic field with
strength $B^{r}=B_{0}$, the deviation from the Landau energy due to the
presence of the other parabola in the effective potential is given by the
following matrix element, which to second-order reads 
\begin{eqnarray}
E_{m}(k\rightarrow \infty ) &=&\left| \left\langle R\right| H\left|
R\right\rangle \left( k\right) \right|  \nonumber \\
&=&\left[ m+\frac{1}{2}\right] -\frac{2^{m-2}}{m!\sqrt{\pi }}\left( 1+\frac{1%
}{p}\right) k^{2m-1}e^{-k^{2}}.  \label{RHR}
\end{eqnarray}
For an electron on the left hand side, i.e. in the smaller magnetic field $%
B^{l}=-B_{0}/p$ region, in the $m^{th}$ Landau level, this results in 
\begin{eqnarray}
E_{m}(k\rightarrow \infty ) &=&\left| \left\langle L\right| H\left|
L\right\rangle \left( k\right) \right|  \nonumber \\
&=&\frac{1}{p}\left[ m+\frac{1}{2}\right] -\frac{2^{m-2}}{m!\sqrt{\pi }}%
\left( 1+\frac{1}{p}\right) k^{2m-1}e^{-k^{2}}.  \label{LHL}
\end{eqnarray}
Also in this case the energy is smaller then the corresponding Landau energy.%
$\ $The downward energy shift decreases for increasing $p$.$\ $

If $p$ is an integer, Landau states on the left and right hand side,
respectively with quantum number $p\cdot m$ and $m$, coincide for $%
k\rightarrow \infty $.$\ $As a consequence these states have an overlap,
which reads to first order 
\begin{eqnarray}
\left\langle L\right| H\left| R\right\rangle &=&\left( -1\right)
^{m+1}2^{m\left( p+1\right) /2}\left( \frac{1}{\left( pm\right) !m!\pi }%
\right) ^{1/2}  \nonumber \\
&&\hspace{1cm}\times p^{pm/2}e^{-k^{2}\left( 1+p\right) /2}k^{m\left(
p+1\right) +1}.  \label{eq:matrix_lhr}
\end{eqnarray}
One can see that for decreasing magnetic field, i.e. increasing $p$, this
function decreases because of the exponential factor.$\ $The electron
wavefunction in the lower magnetic field region is extended over a larger
region, and further away from the other (center at $kp$).$\ $The overlap
therefore decreases with increasing $p$.$\ $As a result of this, the energy
for $k\rightarrow \infty $ and $p>1$ is given by $\left\langle R\right|
H\left| R\right\rangle $ and $\left\langle L\right| H\left| L\right\rangle $.

For $p=1$, we obtain the previous result, but for increasing $p$, the second
order term in Eqs.~(\ref{RHR}) and (\ref{LHL}) becomes more important than
Eq.~(\ref{eq:matrix_lhr}), because of the exponential factor.$\ $The
splitting is lifted, and the main contribution to the negative velocity for $%
k\rightarrow \infty $ arises from Eq.~(\ref{eq:matrix_lhr}) due to the
finite extend of the wavefunction in the other parabola.

As an example we studied numerically the case when a background magnetic
field $B=B_{0}/2$ is applied, i.e. $B^{l}=-B_{0}/2$ and $B^{r}=3B_{0}/2$.$\ $%
As one can see in Fig.~\ref{fig:trajectories}(b), this results in two
parabolas with different minima and confinement strength.$\ $The resulting
spectrum (see Fig.~\ref{fig:spectrum}(b)) is very similar to the one of the
previous case, but unlike the previous symmetrical case, not all states are
twofold degenerate for $k\rightarrow \infty $.$\ $We now obtain two
different sets of Landau states, corresponding to electrons moving in
different magnetic field regions with different strength.$\ $In this case
the second Landau level on the left coincides with the first on the right.$\ 
$The classical picture for the three different regions corresponds to the
one drawn in Fig.~\ref{fig:trajectories}(b), and is also similar to the
previous case, except for the different cyclotron radii.$\ $With this
picture in mind, one can again calculate the classical velocity, which turns
out to be identical to Eq.~(\ref{eq:clasvel1}). From Fig.~\ref{fig:velocity}%
(b) we notice that again we obtain good agreement for $k>0$, but for $k<0$,
there is a large discrepancy.$\ $The negative velocity can also in this case
not be explained classically.

The critical $k$-value $k^{\ast }=\sqrt{2E\left( k^{\ast }\right) }$ for
which snake orbits are classicaly possible are indicated by the parabola in
Fig.~\ref{fig:spectrum}(b).$\ $

\subsection{$B=B_{0}$}

When a background magnetic field $B=B_{0}$ is applied, we obtain the
magnetic barrier studied in Ref.\onlinecite{peeters93}, where the magnetic
field is different from zero only in the region $x>0$, i.e. $B^{l}=0$ and $%
B^{r}=2B_{0}$.$\ $From Fig.~\ref{fig:trajectories}(c) one can see that in
this case the potential is made up of only one parabola and on the left side
it is a constant $k^{2}/2$.$\ $The energy spectrum and corresponding
velocities for this particular case are shown in Fig.~\ref{fig:spectrum}(c)
and ~\ref{fig:velocity}(c), respectively.$\ $We notice that for $%
k\rightarrow \infty $ we again obtain Landau states, which correspond to
bound states on the right hand side of the magnetic edge.$\ $Consistent, as
being a limiting case of the former magnetic field states, i.e. $p=\infty $,
the energy decreases with decreasing $k$ and there is no splitting of the
energy levels.$\ $Thus now we only have states which propagate with negative
velocity to which we cannot assign a classical interpretation.$\ $

Also in this case we can divide up the spectrum into three regions: (A) the
electrons move in closed orbits in the magnetic field region on the right
hand side, (B) electrons are free, propagate forward and are reflected on
the barrier and (C) electrons are free, propagate backward and are reflected
on the magnetic edge.$\ $Notice that for a free electron, the energy is
larger than $k^{2}/2$, since now the electron also propagates in the $x$%
-direction and consequently has an additional kinetic energy $k_{x}^{2}/2$.

Classically, propagating states in the magnetic field region do not exist,
only Landau states do.$\ $The boundary where these classical trajectories
are possible is again given by $k^{\ast }=\sqrt{2E\left( k^{\ast }\right) }$.

\subsection{$B_{0}>B$}

By applying a background magnetic field with strength larger than $B>B_{0}$,
we arrive at the situation where $0<B^{l}<B^{r}$. The magnetic fields on the
left and the right hand side have the same sign, but a different strength,
i.e. $B^{l}=B^{r}/p.$

To obtain the energy in the limits $k\rightarrow \pm \infty $, we again can
approximate the wavefunction as being in a Landau state in the corresponding
magnetic field.$\ $We found 
\begin{eqnarray}
E(k\rightarrow \infty ) &=&\left\langle R\right| H\left| R\right\rangle
\left( k\right)   \nonumber \\
&=&\left[ m+\frac{1}{2}\right]   \nonumber \\
&&-\frac{2^{m-2}}{m!\sqrt{\pi }}\left( 1+\frac{1}{p}\right)
k^{2m-1}e^{-k^{2}},
\end{eqnarray}
for an electron on the right hand side in the $m^{th\text{ }}$Landau state
of a magnetic field with strength $B^{r}=B_{0}$.$\ $For an electron on the
left hand side, in the smaller magnetic field $B^{l}=B_{0}/p$ in the $m^{th}$
Landau level, we have 
\begin{eqnarray}
E(k\rightarrow -\infty ) &=&\left\langle L\right| H\left| L\right\rangle
\left( k\right)   \nonumber \\
&=&\frac{1}{p}\left[ m+\frac{1}{2}\right] +\frac{2^{m-2}}{m!\sqrt{\pi }}%
\left( 1+\frac{1}{p}\right) k^{2m-1}e^{-k^{2}},
\end{eqnarray}
which results in a negative velocity.$\ $

The energy spectrum and the velocity of these eigenstates for the case when $%
B=3B_{0}/2$, i.e. $B^{l}=B_{0}/2$, $B^{r}=5B_{0}/2$, are plotted
respectively in Fig.~\ref{fig:spectrum}(d) and \ref{fig:velocity}(d).$\ $The
center of the orbit is situated on the right side for $k>0$, for $k<0$ it is
on the left side. For $k\rightarrow \pm \infty $, the electrons move in a
homogeneous magnetic field (on the left ($k\rightarrow -\infty $) or right ($%
k\rightarrow +\infty $) hand side of $x=0$), and thus $v_{y}=0$.

From Fig.~\ref{fig:trajectories}(d) one notices that there is only one
minimum in the effective potential because the minima of both parabolas are
now situated on the same side.$\ $The trajectories corresponding with
regions (A), (B), and (C) are depicted in Fig.~\ref{fig:trajectories}(d).$\ $%
The trajectories in region (A') are similar to those in (A) but now for a
magnetic field on the left hand side, i.e. with smaller strength.

Geometrical considerations yield the following classical velocity 
\begin{eqnarray}
v_{y}(k) &=&2v_{\varphi }(k)\sqrt{1-\left[ k/v_{\varphi }(k)\right] ^{2}} 
\nonumber \\
&&\times \{B^{l}\arccos \left[ -k/v_{\varphi }(k)\right]  \nonumber \\
&&\hspace{0.5cm}+B^{r}\arccos \left[ k/v_{\varphi }(k)\right] \}^{-1},
\label{eq:clasvel2}
\end{eqnarray}
which is plotted in Fig.~\ref{fig:velocity}(d) as dotted curves together
with the quantum mechanical group velocity.$\ $One can see that, in contrast
to the previous cases, the negative velocity can be understood as classical
snake orbits, but these snake orbits all run in the same $y$-direction and
now there are no states with $v_{y}>0.$

Notice that the quantum mechanical velocity exhibits a small oscillatory
behaviour on top of a uniform profile. These whiggles can be understood from
the electron distribution over the two parabolas (see Fig.~\ref{fig:whiggles}%
). With increasing $k$, the electron distribution is shifted from the left
parabola to the right one. Due to the wavelike character of this
distribution, the probability for an electron to be in the right parabola
(integrated solid region in the inset of Fig.~\ref{fig:whiggles}) exhibits
whiggles as function of $k$, with $n$ maxima as shown in Fig.~\ref
{fig:whiggles}. Energetically it is favourable for an electron state to have
as much as possible electron probability in the lower potential region.
Consequently, when the electron probability in the lower potential region
attains a maximum, a maximum downward energy shift will be introduced on top
of the overall energy change, and this will result in a maximum in the group
velocity.

\section{Negative velocity state}

Formally, the existence of the quantum mechanical negative velocity state
can be attributed to the fact that shifting two one dimensional potential
wells towards each other results in a significant rearrangement of the
energy levels in the composite potential well.$\ $Because the composed well
is broader, some states, p.e. the ground state, have an energy which is
lower than in each of the individual narrower wells.$\ $In this particular
case, the wave vector $k$ measures the distance of the two wells to each
other, and consequently this energy decrease results in a negative group
velocity $-\partial E/\partial k$. In this section we focus on these
negative group velocity states.$\ $

Since the negative velocity states are present for any background magnetic
field $B$, but can only be understood classically in the situation $B>B_{0}$%
, we will investigate the group velocity for a fixed $k$ value with varying
background magnetic field.$\ $In Fig.~\ref{fig:spectrumb} the spectrum is
plotted as function of the applied background field $B$. We have chosen $%
k=1.5$ because in this case a large negative velocity is obtained for the
lowest level when $B=0$.$\ $Notice that for $B<B_{0}$: (1) almost all levels
decrease in energy with increasing background field; (2) there is an
anti-crossing for $E/E_{0}=0.4+0.458B/B_{0}$ (dotted line). This
anti-crossing occurs when $B/B_{0}=n/(n+1)$, with $n$ the Landau level
index. For this condition some of the Landau levels are degenerate in the
limit $k\rightarrow \infty $ (see Fig.~\ref{fig:spectrum}(b) for the case $%
n=0$); and (3) for $B\rightarrow B_{0}$ the separation between the levels
decreases to zero and a continuous spectrum is obtained with a separate
discrete level at the anti-crossing line. The continuous spectrum for $%
B=B_{0}$ results from the scattered states in the potential of Fig.~\ref
{fig:spectrum}(c), while the discrete state is the bound state in this
potential. For the considered $k$-value, i.e. $k=1.5$, only one bound state
is found for $B=B_{0}$.

The corresponding group velocity $v_{y}=-\partial E/\partial k$ is shown in
Fig.~\ref{fig:velocityb}. Notice that the maximum negative velocity is
obtained near the anti-crossings in the energy spectrum (Fig.~\ref
{fig:spectrumb}). Near $B/B_{0}=n/\left( n+1\right) $ the splitting in the
energy spectrum (see Fig.~\ref{fig:spectrum}) is largest and as a
consequence one of the levels is pushed strongly down in energy and
consequently $v_{y}$ becomes strongly negative. Notice that: (1) every level
has some $B/B_{0}$ region at which $v_{y}<0$, and (2) for $%
B/B_{0}\rightarrow 1$ the velocity $v_{y}\rightarrow 0$, while (3) the
enveloppe of $\left( v_{y}\right) _{\min }$ in Fig.~\ref{fig:velocityb}
reaches for $B/B_{0}$ the $v_{y}<0$ value of the $B=B_{0}$ state. For $%
B>B_{0}$ we have $v_{y}<0$ for all states.

Using expressions (\ref{eq:clasvel1}) and (\ref{eq:clasvel2}), we can also
calculate the classical velocity corresponding to the energy spectrum in
Fig.~\ref{fig:spectrumb}.$\ $This is shown in Fig.~\ref{fig:clasvelb}.$\ $We
notice that for $B/B_{0}<1$, the classical velocity has a similar behaviour
as the quantum mechanical one, except for the anti-crossings and the lack of
negative velocities, which do not have a classical analogon.$\ $These
negative velocities appear suddenly for $B/B_{0}>1$ and exhibit more or less
the same behaviour.$\ $

As was already apparent from the above study a necessary condition for the
existence of the non-classical edge states is the presence of two local
minima in the effective potential. In the limiting case $B=B_{0}$ the second
minima is the limiting case of a flat region in $V_{k}\left( x\right) $ for $%
x<0$. But not all these states have a negative velocity. How can we classify
them?

From Fig.~\ref{fig:spectrum}(a,b) one notices that initially (for rather
small $B$ values) the parabola $E=k^{2}/2$ separates the region where only
states with positive group velocity exist, from the region where also
negative velocity states are present. This is due to the fact that the value
of this parabola equals the barrier height between the two parabolic wells
for the corresponding $k$-value. When the energy exceeds this barrier, the
shape of the wavefunction is not determined anymore by the separate
parabolas, but by the overall composite well width. For decreasing $k$ the
well is squeezed, and thus all the energy levels are pushed upwards,
resulting in a positive group velocity. Although this is not an exact rule
which cannot be extended rigorously throughout the $B<B_{0}$ regime, it
nevertheless provides insight into the $k$-values (or $B$ values) for which
these negative velocities states arise.

Inspection of the wavefunctions shows that there is a feature which marks
the negative velocity states, and which relates indirectly to the presence
of the different potential wells. It turns out that if the wavefunction or
its first derivative exhibits a dip at some $x$ which satisfies $\phi
(x)\phi ^{\prime \prime }(x)>0$ and $\phi ^{\prime }(x)=0$, or $\phi
^{\prime }(x)\phi ^{\prime \prime \prime }(x)>0$ and $\phi ^{\prime \prime
}(x)=0$ and the condition that $\phi (x)\neq 0$, then the state has a
non-classical negative velocity. This is true for every $k$ value, as long
as $B<B_{0}$. This is illustrated in Fig.~\ref{fig:waves_min} where we plot
the wavefunction for $k=1.5$, $n=2$, with background magnetic field $%
B/B_{0}=0.68$ and $0.73$. The above dip in the wavefunction or its
derivative (indicated by the dashed circle in Fig.~\ref{fig:waves_min}) is a
result of the different potential wells, which have their separate influence
on the shape of the wavefunction, and therefore hamper the matching. The
difference in $\phi (x)$ being zero (or not), can be interpreted as a
generalization of matching the individual states in an asymmetric (symmetric
way) when the Landau states are degenerate at $k\rightarrow \infty $.

\section{Time dependent classical Interpretation}

One can make different attempts to link a classical picture to quantum
mechanics.$\ $Often the comparison starts with a schematic classical picture
which is then supported by comparing the quantum mechanical probability
density with the classical one, obtained through calculation of the
classical electron trajectory solving Newtons equation.$\ $For a 1D problem
one can also verify the classical motion by inspection of the velocity
parallel to the edge.\cite{muller92,zwerschke99,gu97,peeters99} For a
cilindrical symmetric problem, the classical electron motion can be inferred
from the magnetic moment or the circular current distribution of the
electron state.\cite{solimany95,sim98,reijniers99,kim99} In this paper, a
quantitative comparison was made by use of a quantum mechanical velocity
parallel to the 1D magnetic field discontinuity. In the following, we will
try a different approach where we will construct time dependent states, and
in doing so we will introduce another feature, i.e. the oscillation
frequency perpendicular to the magnetic edge.

\subsection{$B=0$}

We already mentioned before that the solutions for this kind of problem are
the parabolic cylinder functions $\phi (x)=D_{E-\frac{1}{2}}\left[ \sqrt{2}%
\left( x-k\right) \right] $, matched in such a way that we have symmetric
and anti-symmetric wavefunctions as is shown for $k=2$ in Fig.~\ref
{fig:trajectories}(a).\ At $k=\infty $, these symmetric and antisymmetric
states are twofold degenerate (see the two wavefunctions corresponding to
the solid square in Fig.~\ref{fig:trajectories}(a)). Due to this degeneracy
any linear combination of these states is also an eigenstate. If we take the
following linear combination 
\begin{equation}
\begin{array}{l}
|m_{+}\rangle =\frac{1}{\sqrt{2}}\left( |2m\rangle +|2m+1\rangle \right) ,
\\ 
|m_{-}\rangle =\frac{1}{\sqrt{2}}\left( |2m\rangle -|2m+1\rangle \right) ,
\end{array}
\end{equation}
we arrive at the well known Landau states, i.e. wavefunctions of electrons
located in two different homogeneous magnetic field profiles. One electron
is moving clockwise, while the other is moving counterclockwise.$\ $For
decreasing $k$ this degeneracy is lifted.$\ $Although taking linear
combinations of states with a different energy yields a time dependent
solution, we will extrapolate this picture towards all the other states. We
choose a new orthonormal \-but time dependent\- basis: 
\begin{eqnarray}
|m_{+}\rangle  &=&\left( e^{iE_{2m}t}|2m\rangle +e^{iE_{2m+1}t}|2m+1\rangle
\right) /\sqrt{2}  \nonumber \\
|m_{-}\rangle  &=&\left( e^{iE_{2m}t}|2m\rangle -e^{iE_{2m+1}t}|2m+1\rangle
\right) /\sqrt{2}  \label{eq:time}
\end{eqnarray}
with $E_{m_{+}}=E_{m_{-}}=(E_{2m}+E_{2m+1})/2$. The resulting energy
spectrum is shown in Fig.~\ref{fig:spectrum}(a) by the dashed curves.$\ $The
corresponding velocities are plotted in Fig.~\ref{fig:vel_oscil_0}(a). For
every branch there are two states $|m_{+}\rangle $ and $|m_{-}\rangle $.

With these new quantum states much better agreement is obtained with the
corresponding classical results (dotted curves in Fig.~\ref{fig:vel_oscil_0}%
(a)). Because of the addition of the two eigenstates the negative velocity
almost disappeared.$\ $Only the lowering of the energy, as was mentioned in
the limiting case (i.e. $k\rightarrow \infty $), results in a small negative
velocity, which can't be understood even in this picture.$\ $Also the
boundary which indicates when classical states propagate is in much better
agreement now.$\ $

Since we now have time dependent states, we can calculate a new feature: the
oscillation frequency $\omega _{x}$ in the $x$-direction.$\ $The time
dependent probability densities of the $|m_{+}\rangle $ and $|m_{-}\rangle $
states have the following form: 
\begin{eqnarray}
|\left\langle m_{+}|x\right\rangle (t)|^{2} &=&\frac{1}{2}(|\left\langle
2m|x\right\rangle |^{2}+|\left\langle 2m+1|x\right\rangle |^{2}  \nonumber \\
&&+2\cos [\omega _{x}t]\left\langle 2m|x\right\rangle \left\langle
2m+1|x\right\rangle ),  \nonumber \\
&=&|\left\langle m_{-}|x\right\rangle (t+\pi /\omega _{x})|^{2},
\label{eq:prob_time}
\end{eqnarray}
where $\omega _{x,m}=(E_{2m+1}-E_{2m})/\hbar $ is the quantum mechanical
oscillator frequency in the $x$-direction.$\ $

Classically we can calculate this frequency $\omega _{x}$ again, using
simple geometrical considerations, which results in 
\begin{equation}
\omega _{x}(k)=\frac{\pi }{2\arccos (-k/v_{\varphi }(k))}.
\end{equation}
Both results are plotted in Fig.~\ref{fig:vel_oscil_0}(b), and we obtain
reasonably good agreement between the quantum (solid curves) and classical
(dotted curves) results.$\ $Notice that for $\left| k\right| >k^{\ast }$,$\ $%
classically $\omega _{x}=0$, which means that the electron does not
oscillate between the two different magnetic field regions (i.e. it is not a
snake orbit state), but it oscillates in a homogeneous magnetic field and
consequently we obtain the time independent eigenstates corresponding to the
Landau levels.

Of course this approach is only useful if proper linear combinations are
possible.$\ $Unfortunately this is not the case when a background magnetic
field $B\leq B_{0}$ is applied.$\ $

\subsection{$B>B_{0}$}

The above approach is also fruitful in the case when $0<B^{l}<B^{r}$. We can
again add adjacent levels, two by two, similar as described before.$\ $We
can repeat exactly as was done before, and we arrive again at the time
dependent states of Eq.~(\ref{eq:time}).$\ $The energy spectrum of these
states when $B=3B_{0}/2$ is shown in Fig.~\ref{fig:spectrum}(d), by dashed
curves.$\ $From Fig.~\ref{fig:vel_oscil_15}(a), we notice that the classical
velocity is in better agreement then before, since the amplitude of the
whiggles is lowered, due to the summation.$\ $The quantum mechanical
oscillation frequency in the $x$-direction is again given by $\omega
_{x,m}=(E_{2m+1}-E_{2m})$ and plotted in Fig.~\ref{fig:vel_oscil_15}(b).$\ $%
We notice that since there are no degenerate states, we always have
oscillating electrons.$\ $For $k\rightarrow \infty $ the electron oscillates
with frequency $\omega _{x}=2.5\omega _{c}$, while for $k\rightarrow -\infty 
$ the electron oscillates with frequency $\omega _{x}=0.5\omega _{c}$, i.e.
the electrons circle around in their seperate homogeneous magnetic fields.$\ 
$This can also be seen from the classical oscillation frequency in the $x$%
-direction, which is given by 
\begin{equation}
\omega _{x}(k)=\pi \left[ \frac{\arccos (k/v_{\varphi }(k))}{B^{l}}+\frac{%
\arccos (-k/v_{\varphi }(k))}{B^{r}}\right] ^{-1}.
\end{equation}
Notice that also here whiggles in $v_{y}$ are present (see Fig.~\ref
{fig:vel_oscil_15}(b)) which are not present in the classical results.$\ $It
is clear that proper linear combinations can always be made, as long as$\
B>B_{0}$.

\section{Conclusions}

We studied the electron states near discontinuities in the magnetic field.
Different 1D magnetic field profiles, i.e. steps, were considered. The
quantum mechanical energy spectrum was obtained and the group velocity of
the states was calculated. Their corresponding classical orbits were found
and the propagating states which are located at the magnetic field
discontinuity correspond to snake orbits. Quantum mechanical magnetic edge
states were found which move along the magnetic field step in opposite
direction to the classical snake orbits and which cannot be understood
classically. We were able to construct non stationary quantum mechanical
states which closely approximate the classical solution for the symmetrical
case $B^{l}=-B^{r}$ and for the more general case $B^{r}>B^{l}>B_{0}$.

\acknowledgments{This work was partially supported by the Inter-university Micro-Electronics Center (IMEC, Leuven),
the Flemish Science Foundation (FWO-Vl), BOF-GOA and the IUAP-IV.  J.R. was
supported by ``het Vlaams Instituut voor de bevordering van het Wetenschappelijk \& 
Technologisch Onderzoek in de Industrie" (IWT) and F. M. P. is a research director
with the FWO-Vl.  We acknowledge fruitful discussions with A. Matulis, P. Vasilopoulos, S. Badalian and J. A. Tyszynski.
}

\begin{figure}[tbp]
\caption{The effective potential (left part of the figure) for $k=-2$
(dotted curve)\ and $k=2$ (solid curve) for the different magnetic field
profiles indicated at the top of Fig. \ref{fig:spectrum}. The wavefunctions
corresponding to the states in Fig. \ref{fig:spectrum}, as indicated by the
corresponding symbols, are also shown. Schematic representation of the
classical electron trajectories (right part of the figure) corresponding to
the different regions indicated in Fig. \ref{fig:spectrum}. }
\label{fig:trajectories}
\end{figure}
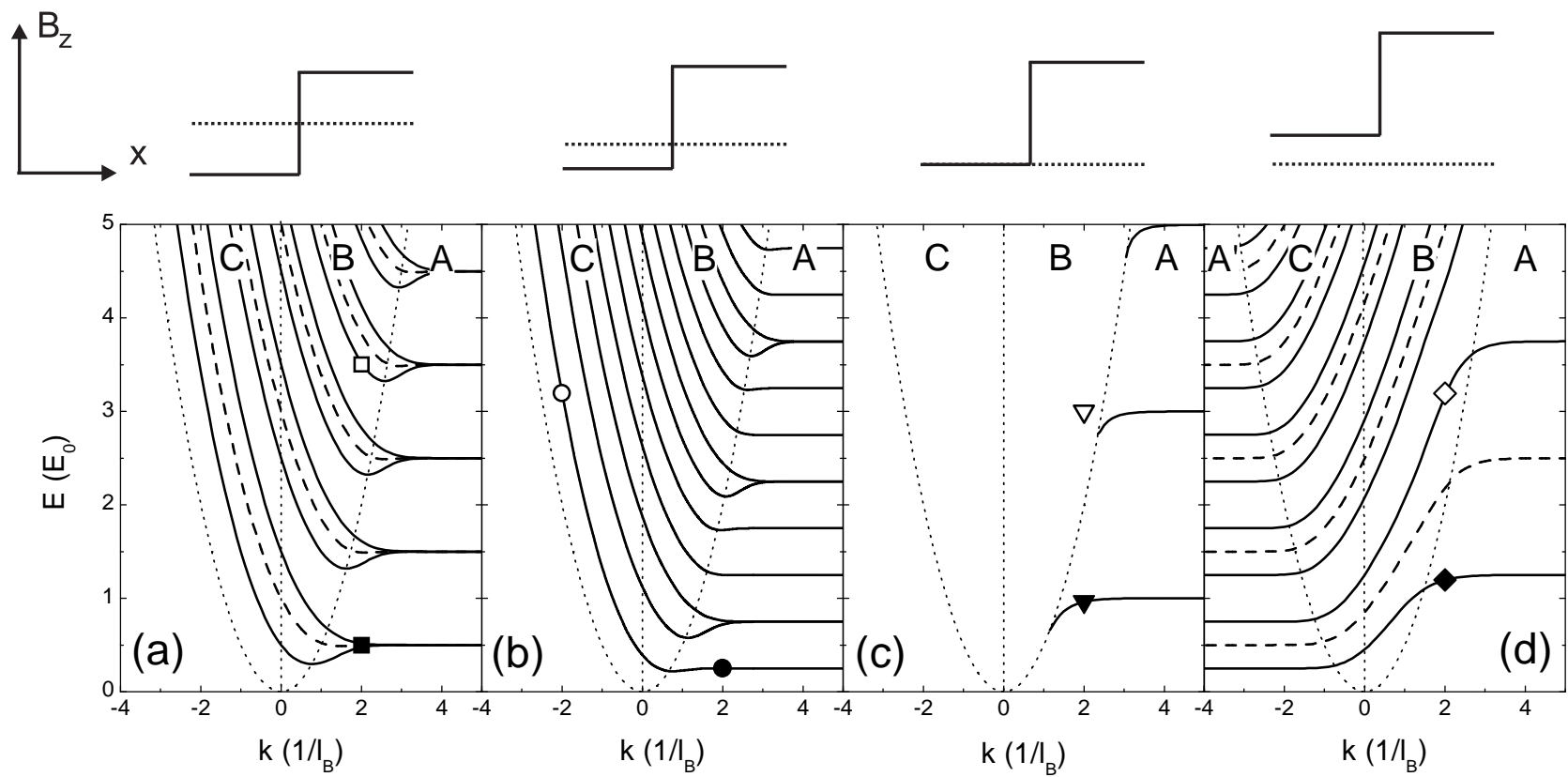
\begin{figure}[tbp]
\caption{The energy spectra for different magnetic field profiles ($B^{l}$,$%
B^{r}$) shown on top of the figure: (a) ($-1,1$), (b) ($-0.5,1.5$), (c) ($0,2
$), (d) ($0.5,2.5$). The dotted curves mark the different classical regions.
The dashed curves in (a) and (b) are the average energy of two adjacent
levels. \ The symbols correspond to the electron wavefunctions plotted in
Fig. \ref{fig:trajectories}.}
\label{fig:spectrum}
\end{figure}
\begin{figure}[tbp]
\caption{The group velocity in the $y$-direction (parallel to the edge) of
the 6 lowest branches of Fig.~\ref{fig:velocity}. \ The dotted curves
correspond to the velocity calculated for the classical snake orbits.}
\label{fig:velocity}
\end{figure}
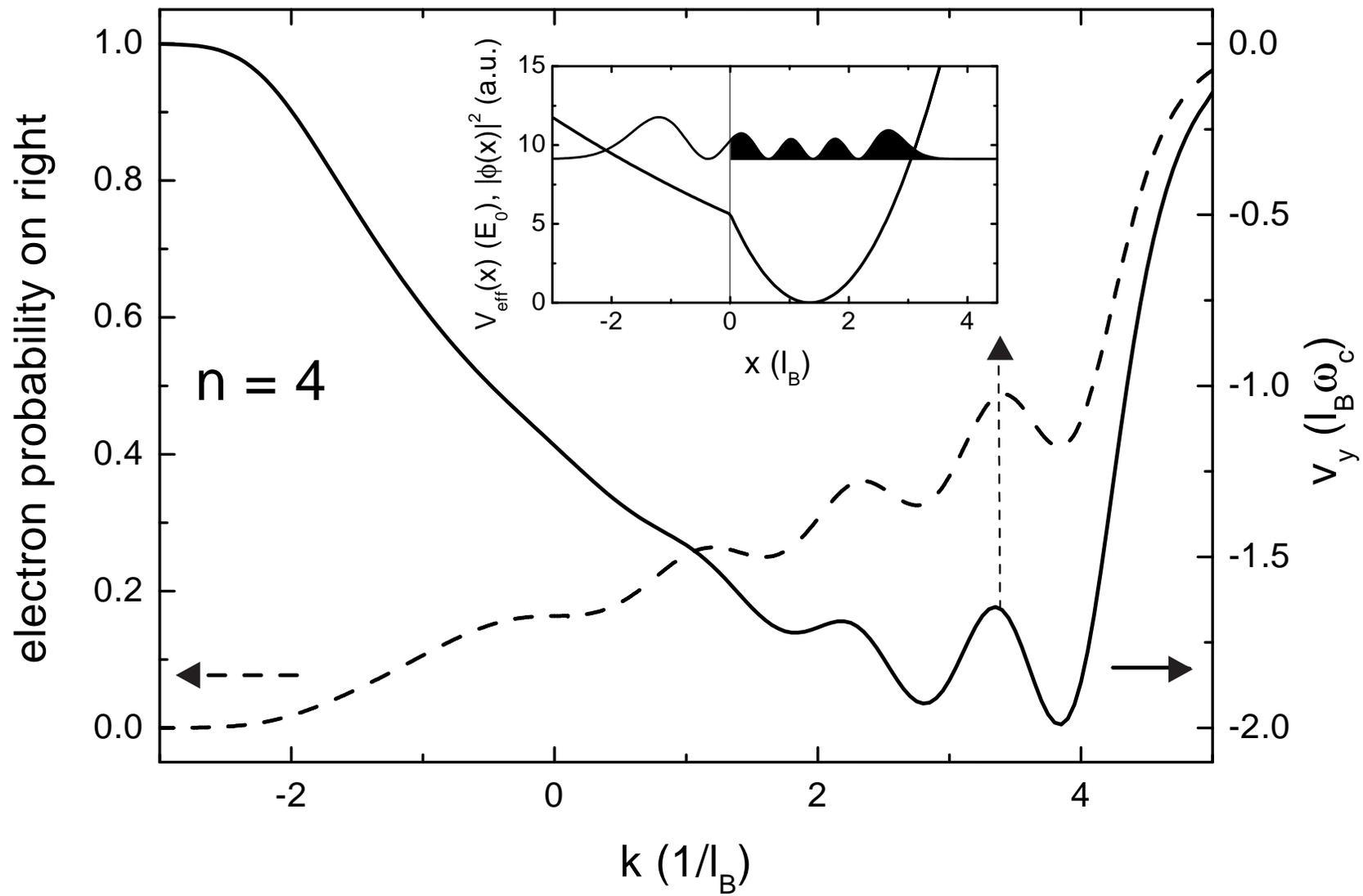
\begin{figure}[tbp]
\caption{The velocity for the $n=4$ state of Fig.~\ref{fig:velocity}(d), and
the electron probability for the electron to be in the right part of the
parabola as function of $k$. \ The inset shows the effective potential and
the electron probability at the local maximum $kl_{B}=3.4$.}
\label{fig:whiggles}
\end{figure}
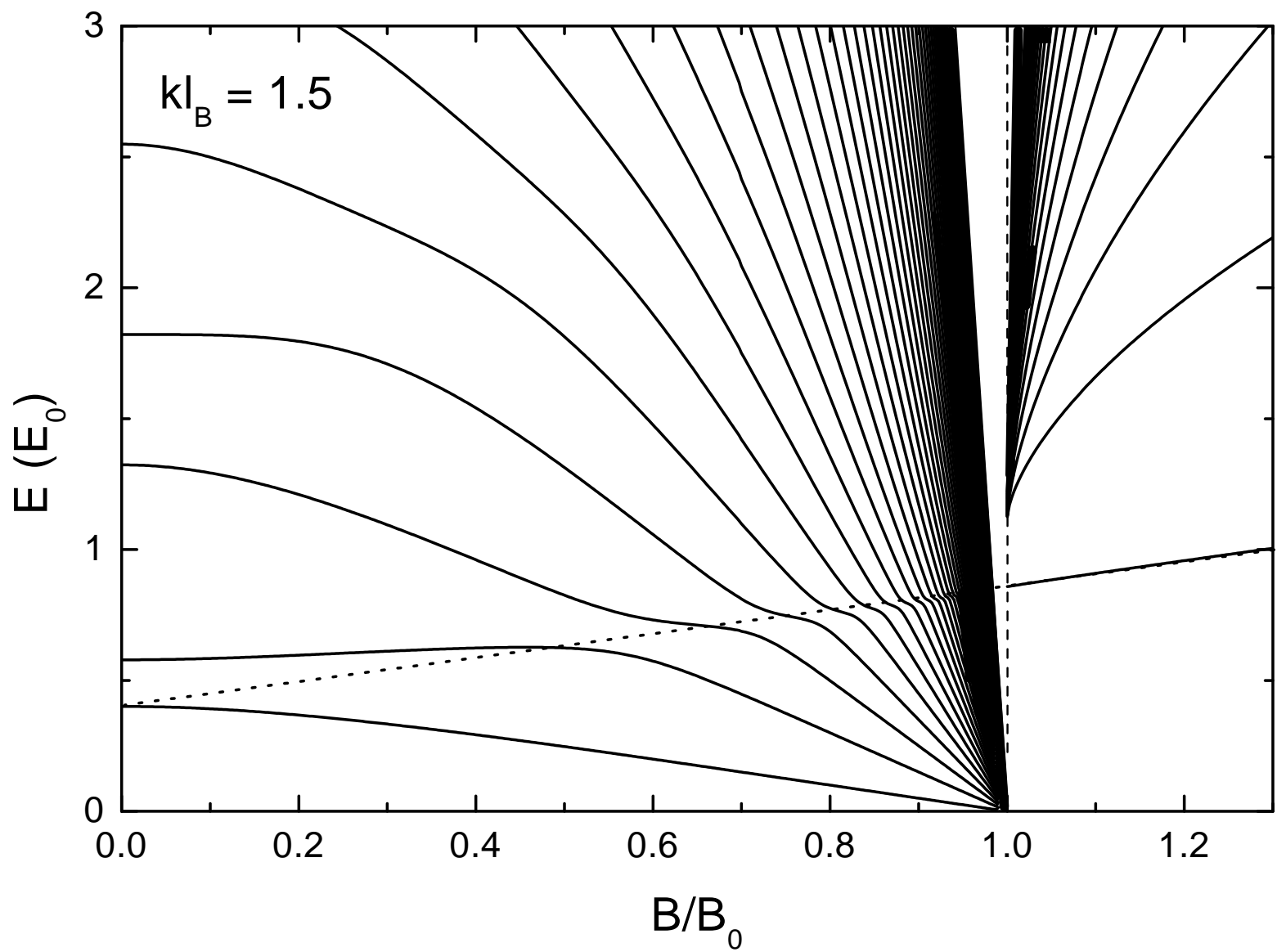
\begin{figure}[tbp]
\caption{The energy spectrum at $kl_{B}=1.5$ for the 40 lowest states as
function of the applied background field $B/B_{0}$. \ }
\label{fig:spectrumb}
\end{figure}
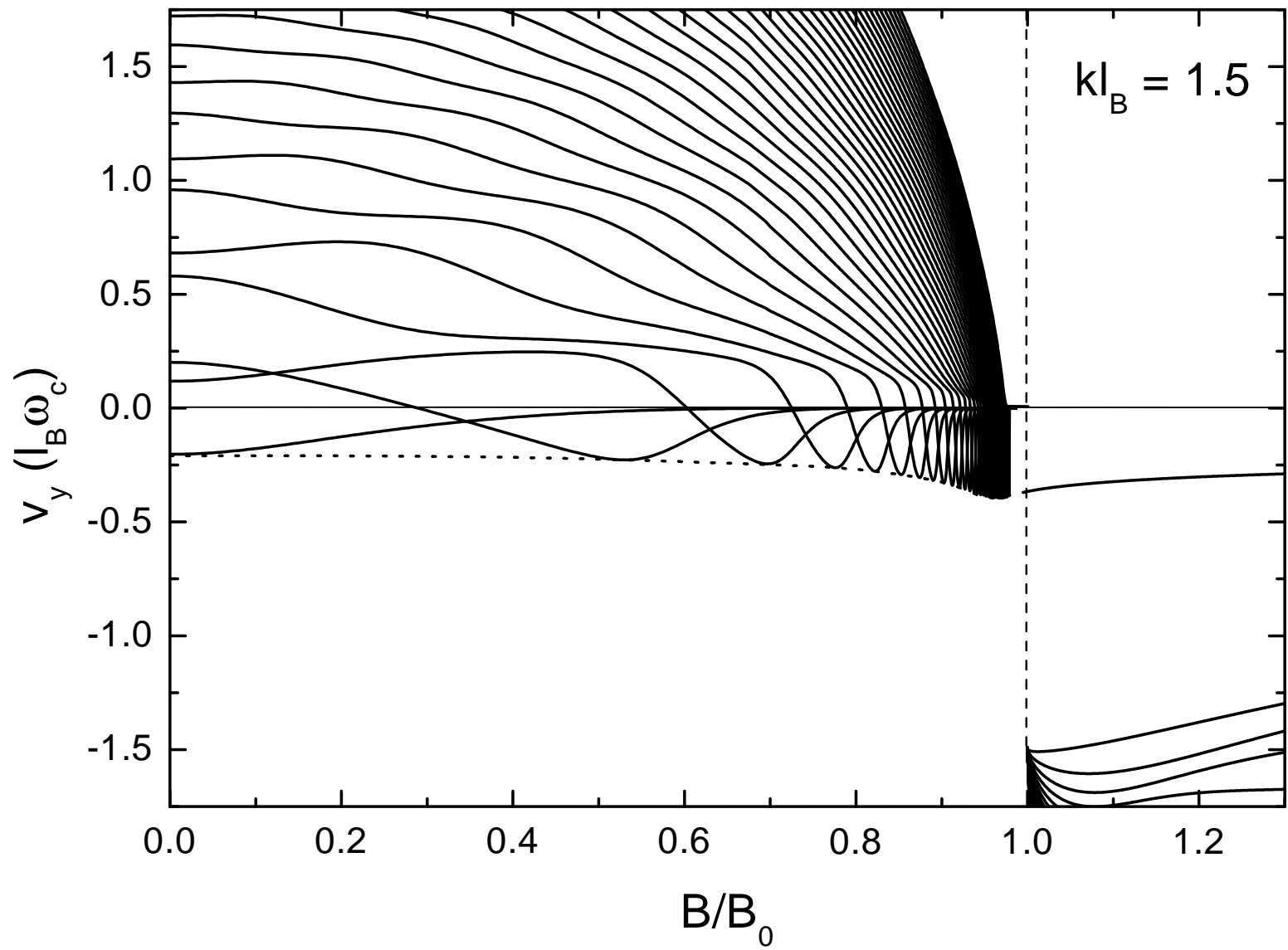
\begin{figure}[tbp]
\caption{The group velocity $v_{y}$ as function of $B/B_{0}$, corresponding
to the electron states of Fig.~\ref{fig:spectrumb}.}
\label{fig:velocityb}
\end{figure}
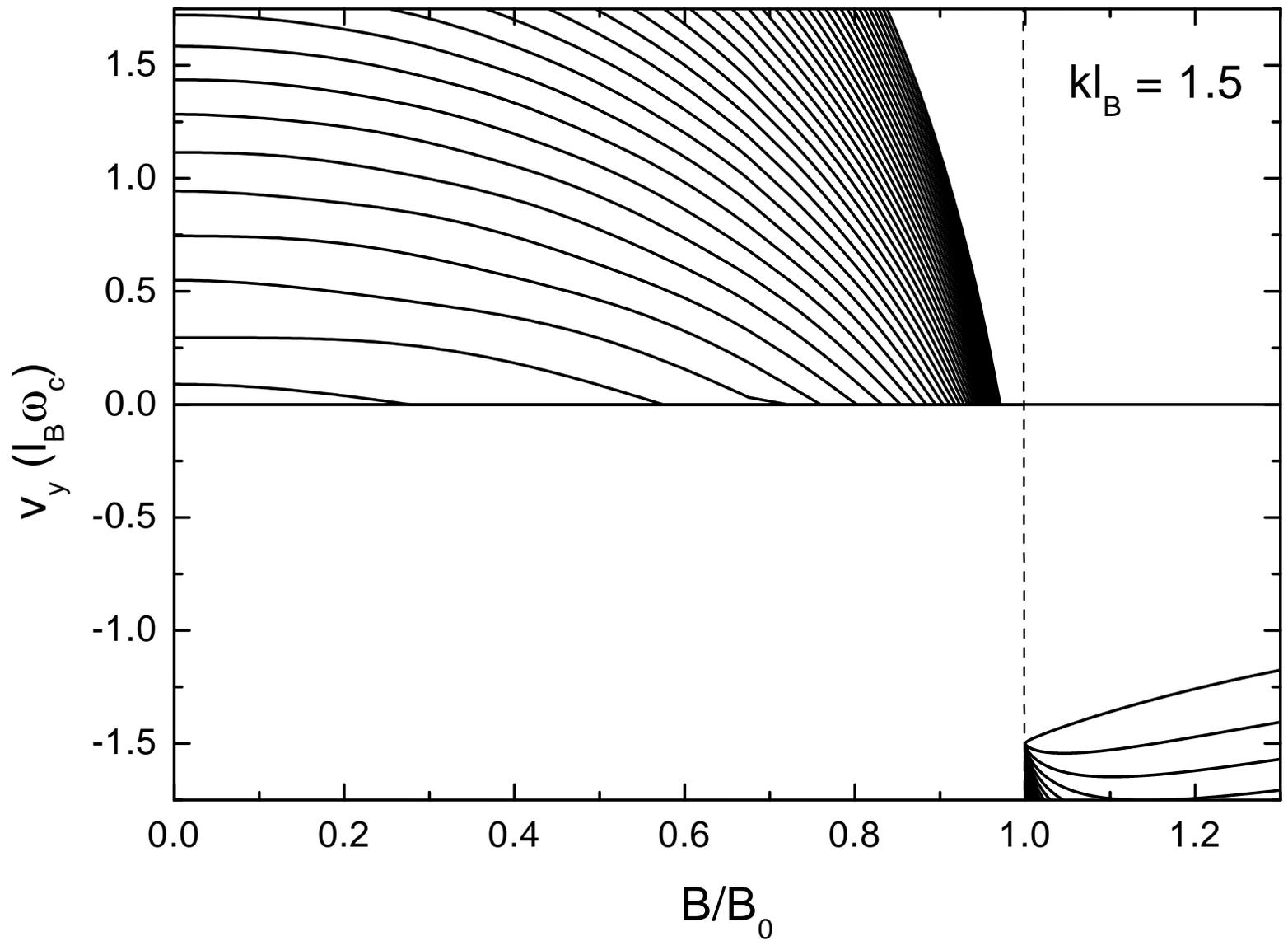
\begin{figure}[tbp]
\caption{The classical velocity in the $y$-direction as function of $B/B_{0}$%
, corresponding to the electron states of Fig.~\ref{fig:spectrumb}.}
\label{fig:clasvelb}
\end{figure}
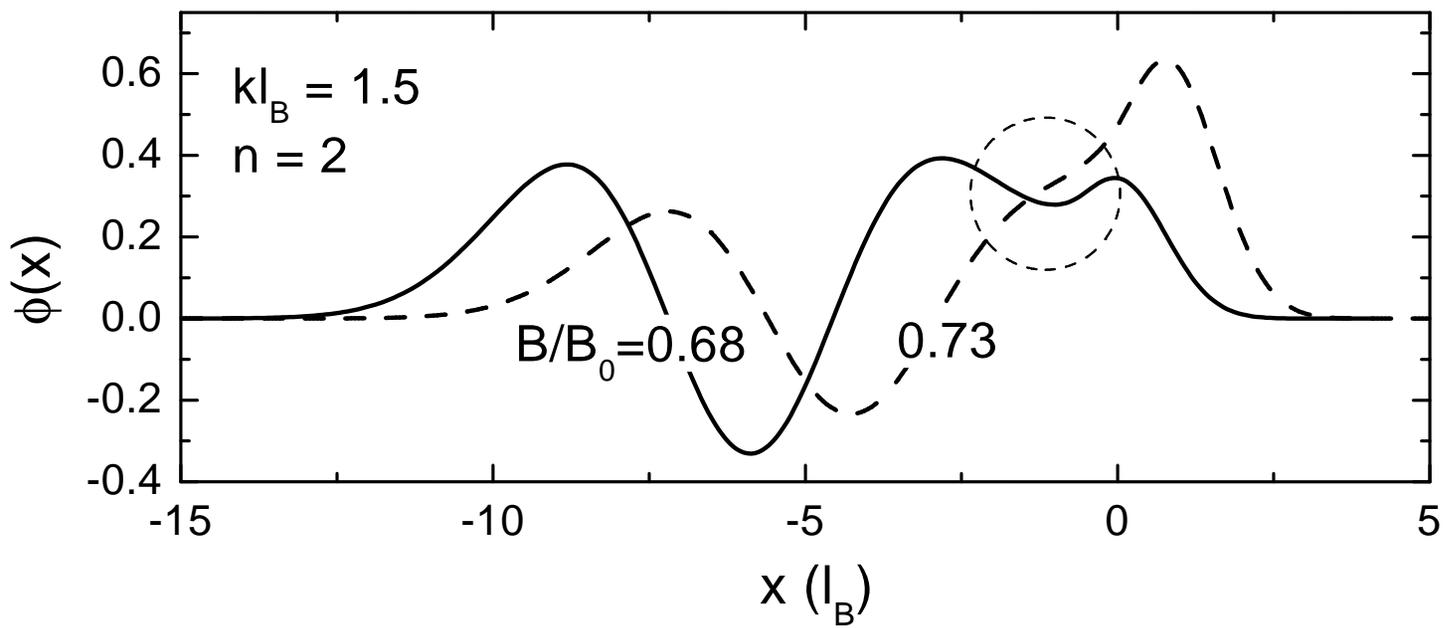
\begin{figure}[tbp]
\caption{The wavefunctions for $kl_{B}=1.5$, $n=2$ for two negative velocity
states (see Fig.~\ref{fig:velocityb}) with background magnetic field $%
B/B_{0}=0.68$; $0.73$.}
\label{fig:waves_min}
\end{figure}
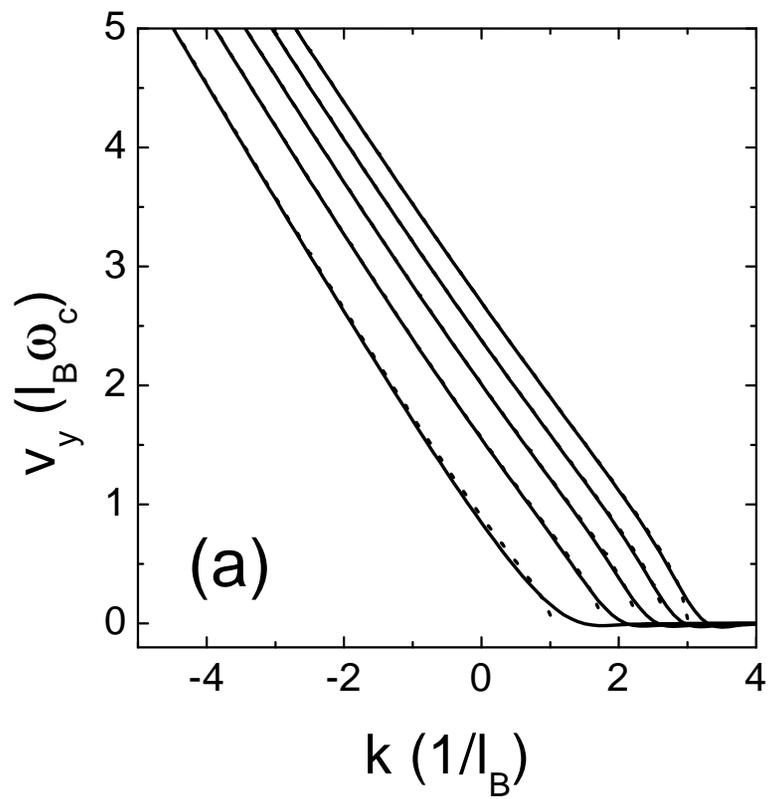
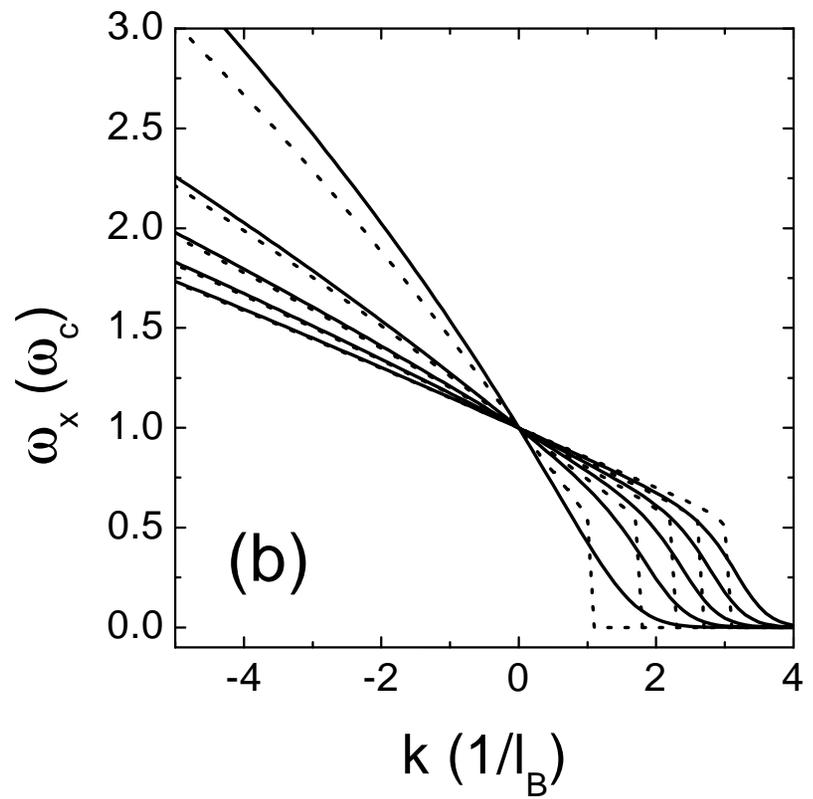
\begin{figure}[tbp]
\caption{(a) The quantum mechanical velocity in the $y$-direction (solid
curves) of the time dependent states for the case $B^{l}=-B^{r}=-1$ and the
velocity obtained classically (dotted curves). (b) The same as in (a) but
now for the oscillation frequency in the $x$-direction.}
\label{fig:vel_oscil_0}
\end{figure}
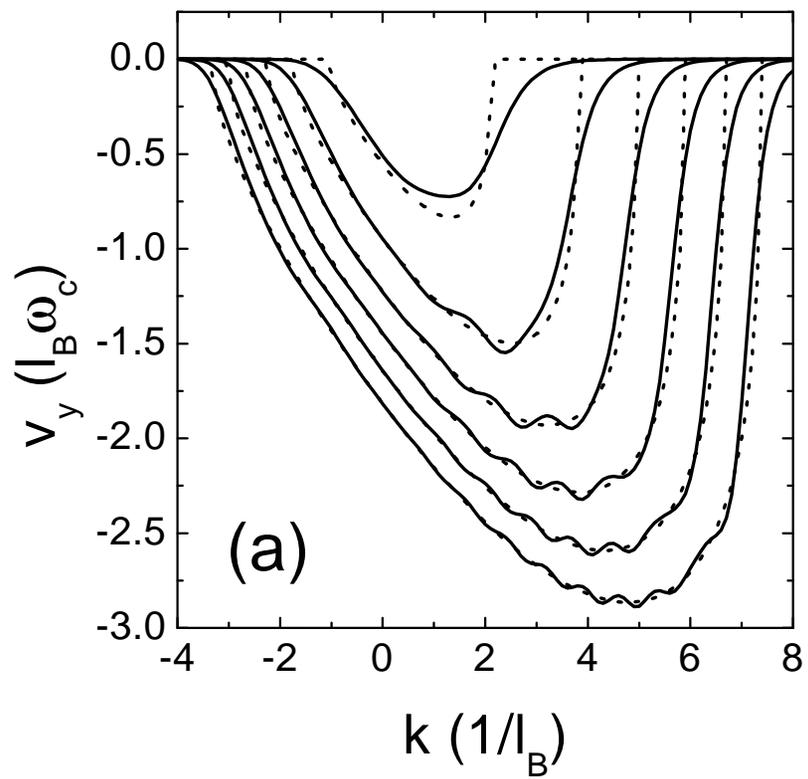
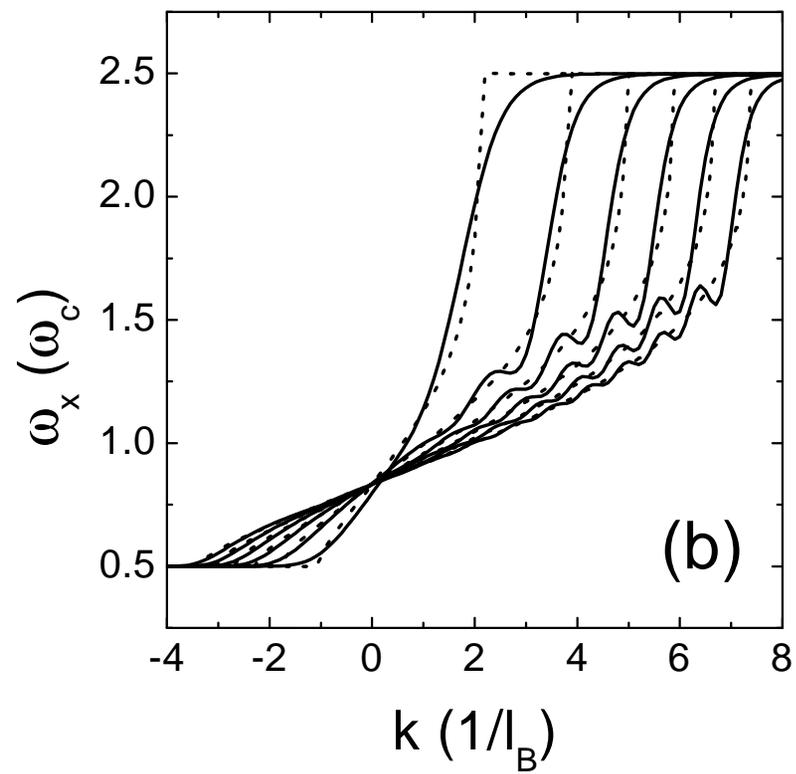
\begin{figure}[tbp]
\caption{The same as in Fig.~\ref{fig:vel_oscil_0}, but now for the magnetic
field profile with $(B^{l},B^{r})=(0.5,2.5)$.}
\label{fig:vel_oscil_15}
\end{figure}

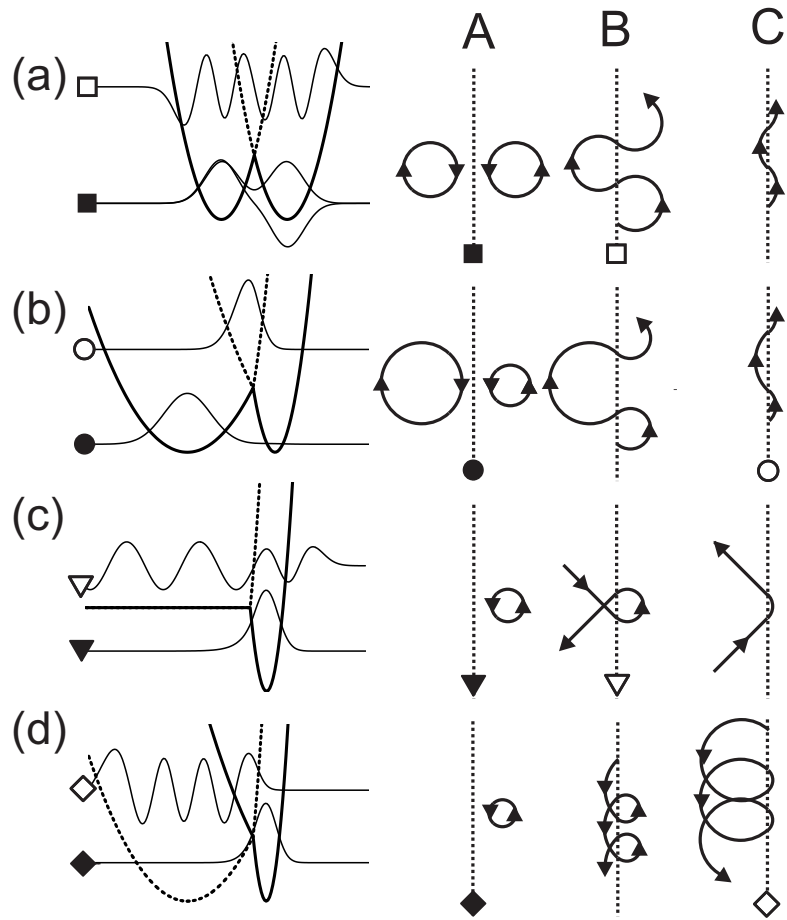
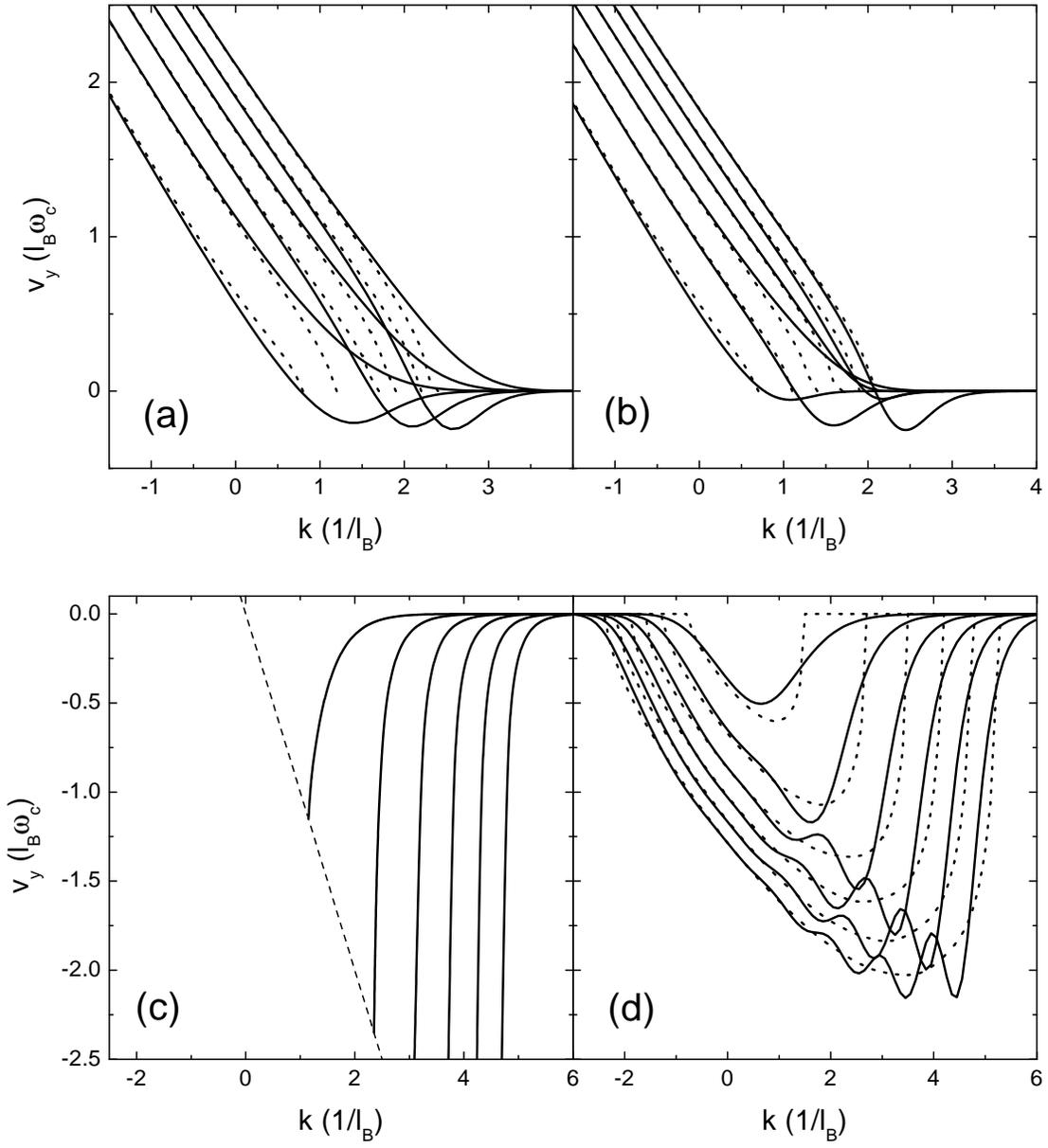
\end{document}